OG 2.4.31

# An All-Sky Search for Steady VHE Gamma-Ray Sources

K. Wang[1] for the Milagro Collaboration

[1] *Department of Physics, University of California, Riverside, CA 92521, USA*

**Abstract**

The Milagrito water Cherenkov detector in the Jemez Mountains near Los Alamos, New Mexico took data from February 1997 to April 1998. Milagrito served as a prototype for the larger Milagro detector, which has just begun operations. Milagrito was the first large-aperture gamma-ray detector with sensitivity to gamma rays below 1 TeV. We report here on a search for steady emission from point sources over most of the northern sky using data from Milagrito.

## 1 Introduction:

The discovery of TeV γ-ray sources in the universe has greatly enriched our knowledge of the astrophysics of particle acceleration. TeV (very-high energy, VHE) gamma rays have been observed by air Cherenkov telescopes (ACT) from at least three galactic and three extragalactic sources (see, for example, Ong, 1998; Hoffman et al., 1999). In addition, ACTs have searched for VHE emission from a number of other sources including some supernova remnants and other blazars. These searches have generally involved exposures of only a few hours to a particular source so they may have missed highly variable sources such as blazars. Because an ACT is a pointed instrument with a field of view of only several millisteradians, there has been no all-sky search for VHE sources. There have been several all-sky searches at higher energies using scintillator arrays with negative results (Alexandreas et al., 1991; McKay et al., 1993).

Milagrito was built and operated as a prototype for the Milagro detector (McCullough *et al.,* 1999). Milagrito, which took data from February 1997 to April 1998, had 228 photomultiplier tubes (PMTs) on a $3 \times 3$ m$^2$ grid under 0.9 m of water. The properties of Milagrito are discussed elsewhere in these Proceedings (Westerhoff *et al.*, 1999). Nearly $9 \times 10^9$ events were recorded from Milagrito. Milagrito was the first air-shower array with sensitivity to gamma rays below 1 TeV. Because Milagrito had a large field of view (>1 sr) and operated all the time, it can be used to search for steady VHE sources anywhere in the northern sky. The sky coverage of Milagrito is illustrated in Figure 1.

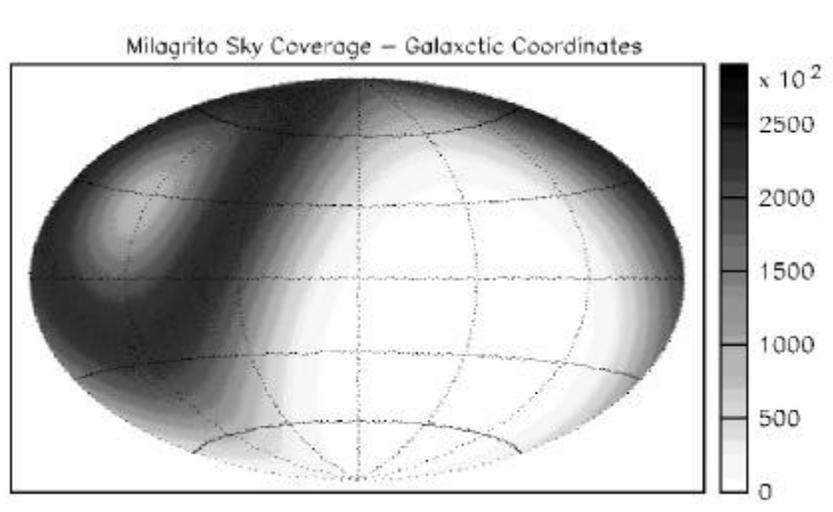

Figure 1. The density of events from Milagrito (in arbitrary units) plotted in galactic coordinates.

## 2   Technique:

The all-sky search for steady TeV sources is performed by dividing the sky into a grid of non-overlapping bins in celestial coordinates, right ascension (RA) and declination ($\delta$). To search for VHE emission from point sources, the number of events in each bin for the entire data set is tabulated and compared with the expected number of background events from cosmic rays. A description of the background estimation method has been discussed in Alexandreas (1993). Because a point source could lie near the edge of a bin, additional searches are made with grids offset in RA, in $\delta$, and in both RA and $\delta$. The bin size is chosen to maximize the significance of a signal in that bin. For a Gaussian angular resolution, this bin is $\pm 1.4\sigma°$ in declination and $\pm 1.4\sigma°/\cos(\delta)$ in right ascension, where $\sigma$ is the rms angular resolution.

The shower direction is calculated from the relative times at which the PMTs are struck after correcting for the effects of electronic slewing, sampling of particles in the shower front, and curvature of the shower front. After making these corrections, the direction of the shower plane can be determined with a least squares fit using the measured times and positions from the PMTs; in reality, some modifications to a straightforward least squares fit are needed to account for the tail of late light due to low-energy particles that tend to trail the shower front and nearly horizontal light in the water from the large Cerenkov angle and from scattering of particles and light in the water. Baffles have been installed around the PMTs in Milagro to block the horizontal light.

Detailed studies of the Milagrito angular resolution have been performed using both data and Monte Carlo simulations. The uncertainty in the reconstructed shower direction can be studied with data using DELEO, which is obtained by fitting each shower with two independent, interleaved portions of the detector (the detector is divided as light and dark squares of a checkerboard) and computing the difference in the fit space angles. DELEO is not sensitive to certain systematic errors such as those due to core location errors. In the absence of these systematic effects, DELEO should be about twice the overall angular resolution (Alexandreas *et al.*, 1992). Figure 2 shows the median DELEO for Milagrito data as a function of the minimum number of PMTs in the fit (NFIT). This shows that the angular resolution is a strong function of NFIT.

The angular resolution can also be obtained from a study of the observed shadow of the moon, after correcting for the bending of the cosmic rays in the earth's magnetic field (Wascko *et al.*, 1999). Unlike DELEO, this technique can reveal systematic pointing errors.

Both of these techniques only address the angular resolution for hadron-induced showers. Monte Carlo simulations are used to compare the expected DELEO distribution for cosmic ray showers to the measured DELEO distribution, and to compare the overall angular resolution for cosmic-ray and photon-induced events. Based on this information, a bin size of $\pm 1.1°$ in declination and $\pm(1.1/\cos(\delta))°$ in right ascension has been chosen for the all-sky search.

In addition, an unbinned search for sources has been performed based on the wavelet formalism. The spatial scale size of this analysis can be varied allowing both point and extended sources to be identified.

Results of the search for steady VHE emission from point sources from these data will be presented, and the sensitivities of the two search techniques will be given. These searches would detect a point source at $\delta = 36°$ (*i. e.* passing overhead) with a steady flux above 1 TeV larger than about $10^{-6}$ m$^{-2}$ s$^{-1}$, assuming a spectral shape similar to that of the Crab. The sensitivity decreases for sources at other declinations; the flux required for detection of a source at $\delta = 25°$ is about 15% larger.

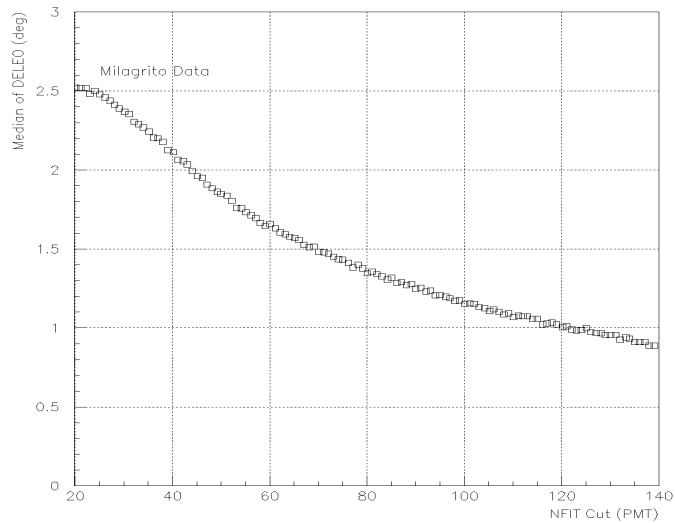

Figure 2. The median DELEO vs. the minimum number of PMTs used in the fit (NFIT) for Milagrito data.

This work was supported in part by the National Science Foundation, The U. S. Department of Energy Office of High Energy Physics, The U. S. Department of Energy Office of Nuclear Physics, Los Alamos National Laboratory, the University of California, the Institute of Geophysics and Planetary Physics, The Research Corporation, and the California Space Institute.